# HybridRanker: Integrating network structure and disease knowledge to prioritize cancer candidate genes

Razieh Abdollahi[a], Zahra Razaghi-Moghadam[a*], Sama Goliaei[b] and Morteza Ebrahimi[a]

[a] *Faculty of New Sciences and Technology, University of Tehran, Tehran, Iran.*
[b] *University of Tehran, Tehran, Iran.*
[*] *Corresponding author: Dr. Zahra Razaghi-Moghadam, email:razzaghi@ut.ac.ir*

{*s.r_abdollahi,razzaghi,sgoliaei, mo.ebrahimi}@ut.ac.ir*}

One of the notable fields in studying the genetics of cancer is disease gene identification which affects disease treatment and drug discovery. Many researches have been done in this field. Genome-wide association studies (GWAS) are one of them that focus on the identification of diseases-susceptible loci on chromosomes. Recently, computational approaches, known as gene prioritization methods have been used to identify candidate disease genes. Gene prioritization methods integrate several data sources to discover and prioritize the most probable candidate disease genes. In this paper, we propose a prioritization method, called HybridRanker which is a network-based technique and it also uses experimental data to identify candidate cancer genes. We apply our proposed method on colorectal cancer data. It is notable to say that in HybridRanker, for considering both local and global network information of a protein-protein interaction network, different algorithms such as shortest-path, random walk with restart and network propagation are exploited. By using these algorithms, initial scores are given to genes within the network. Furthermore, by looking through diseases with similar symptoms and also comorbid diseases and by extracting their causing genes, the gene scores are recalculated. We also use gene-phenotype relations for an additional scoring of the candidate genes. Our method is validated and compared with other prioritization methods in leave one-out cross-validation and the comparison results show the better performance of the HybridRanker.

1. **Introduction**

   Cancer is one of the leading causes of death in the world. The early diagnosis of cancer may prevent the spread of it to other parts of the body and reduce early death risk. So, discovering cancer causing genes is a crucial topic in studying the genetics of cancer. Linkage analysis is an experimental approach to explore diseases-susceptible loci on chromosomes. But, experimental explorations are both time consuming and expensive and are not always feasible [1]. More feasible approaches for this problem are computational ones, known as gene prioritization methods. Prioritization methods leverage experimentally verified disease genes to identify new candidate disease genes for further analysis. To obtain more reliable results, these methods usually integrate different information from several available sources, such as sequence properties, functional annotations, transcriptomic data, molecular pathways, protein-protein interactions and literatures [2-5].

   A number of studies collected biomedical keywords of known disease genes and used text-mining techniques to rank candidate genes based on the text similarity [6, 7]. In some other researches, basic knowledge such as Gene Ontology (GO) or sequence similarity were gathered to construct similarity profiling and they scored candidate genes using this similarity profiling [2, 4, 8].

   Under the assumption that the genes cause the same disease, tend to have physical interactions with each other or participate in the same functional pathways, protein-protein interaction (PPI) networks are important sources for gene prioritization problem and their integration with other data helps to identify more reliable cancer-causing genes.

   Early network-based prioritization methods like molecular triangulation [9], shortest-path (SP) [10] and direct neighbors [11] focus on network local information which lead to vulnerable results. To improve the performance of these prioritization methods, other techniques such as random walks with restart (RWR) [12] and network propagation (NP) [13] utilized network global information. We remark that one of the deficiencies of the later approaches is ignoring genes with few connections in the PPI network [14].

   To overcome the disadvantages of aforementioned methods, we propose a new approach, called HybridRanker, which benefits from both local and global information of the network. Additionally, in developing our method we use the assumption that disease comorbidity and the similarities in symptoms and phenotypes are arisen from common molecular mechanisms [15-18]. Because of all mentioned reasons, we composed RWR, SP and NP algorithms in HybridRanker and also exploited genetic data of diseases with similar symptoms and similar phenotypes, and comorbid diseases as well. All these made our approach a hybrid method based on both network structure and biological knowledge.

As colorectal cancer (CRC) is the third most common type of cancer and one of the major cause of mortality in the world [19, 20], we applied HybridRanker on CRC data. For evaluating the performance of the HybridRanker, it was validated and compared with other prioritization methods in leave one-out cross-validation analysis. The results have shown the better performance of HybridRanker.

## 2. Material and Methods

### 2.1. *Disease genes prioritization*

Network topology is a crucial issue in gene prioritization methods. In a given network, local information of a node relates to the topological neighborhood of that node, while global information characterizes the role of that node within the whole network [14]. In HybridRanker, we utilized both local and global information of the network to improve the performance of previous methods. We applied SP algorithm to extract local information within the network and for obtaining global information of the network, RWR and NP algorithms were used. Initially we scored all proteins within the network by these three algorithms. Furthermore, we exploited the information about disease similarity, comorbidity and phenotype similarity in our method. These steps are described in the following sections:

*RWR algorithm*

By using RWR algorithm all proteins within a given network were ranked based on their proximity to seed genes. Random walks in this algorithm started at seed genes and in each iteration, with a given probability, went to the neighbors of the current nodes or restarted at seed genes. The final ranking was computed by the steady state probability vector of these random walks. In each iteration, the probability vector was calculated by:

$$P^{t+1} = (1-r)WP^t + rP^0 \qquad (1)$$

where $r$ was set to be the restart probability, $W$ was the weighted adjacency matrix of the given network with normalized columns. $P^0$ was the vector of primary probabilities of nodes and $P^t$ contained the probabilities of being at each node in the $t$-th iteration.

Initially, in our method, equal probabilities were assigned to all seed genes and other genes were given zero probability. We tested RWR with different values of $r$ and the best accuracy was obtained at $r = 0.15$. The steady state was achieved by performing iterations until the difference between two consecutive probability vector ($P^t$ and $P^{t+1}$) became less than $10^{-6}$ (the difference is measured by $L_1$ norm) [12].

*Network propagation algorithm (NP)*

NP algorithm is very similar to the RWR algorithm and it also uses network flow propagation to prioritize genes. In RWR, the adjacency matrix is also normalized but in a different way. In NP algorithm both input and output flows of nodes are normalized, in the other words, rows are also normalized in the adjacency matrix [13]. NP algorithm was used to give a rank to each protein in the PPI network.

*SP algorithm*

In addition to previous steps, proteins within the network were ranked with respect to their shortest-path distance to seed ones. The shortest-path distance between seed genes and all other genes were calculated by Dijkstra [21] algorithm. For each non-seed gene, its minimum shortest-path distance to the seed genes was used for its ranking. It was assumed that being close to a seed gene would be a good evidence for a gene to be associated with that disease [22].

*Disease similarity information*

To explore disease with similar symptoms, Human Symptoms Disease Network (HSDN) [23] was used. There were 36 diseases in HSDN which shared similar symptoms with CRC. The causing genes of these diseases were extracted from DisGeNET [24] database. It is possible that diseases with similar symptoms arise because of sharing common disease genes. To consider the similarity of disease symptoms in scoring causing genes, initially all genes were given "zero" score. Iteratively, we went through each similar disease with CRC and the score of its causing genes were added by one. Finally, the score of a gene showed the number of similar (in term of symptoms) diseases it was associated with. For instance, if a gene was a common causing gene of four similar diseases, its final score would be four. The higher score a gene is given, the more likely this gene is CRC-related. So, genes with higher scores were ranked better.

We selected both the top ten and the top five most similar diseases for scoring genes and the better accuracy was achieved when the top ten similar diseases were used. These ten diseases are listed in Table 1.

Table 1. The ten most similar diseases to CRC based on their symptoms (extracted from HSDN).

| Disease | Symptom similarity score |
|---|---|
| Fibrous Dysplasia, Polyostotic | 0.971 |
| Port-Wine Stain | 0.749 |
| HIV Infections | 0.746 |
| Breast Neoplasms, Male | 0.599 |
| Ileitis | 0.526 |
| Esophageal Neoplasms | 0.443 |
| Neoplastic Syndromes, Hereditary | 0.422 |
| Leukemia, Myeloid, Acute | 0.421 |
| Microcephaly | 0.295 |
| Hyperlipidemias | 0.380 |

*Disease comorbidity information*

We explored MalaCards [25] database for comorbid diseases with CRC (in April 2015) and then their causing genes were extracted from DisGeNET database (six comorbid diseases with CRC are shown in Table 2). Finally, to exploit comorbidity information in HybridRanker, genes were scored in the similar way as stated in the previous step.

Table 2. Comorbid diseases with CRC (extracted from MalaCards database).

| Disease name |
|---|
| Adenoma |
| Breast Cancer |
| HIV Infections |
| Familial adenomatous polyposis |
| Lunch-Syndrom |
| Ovary Cancer |

*Gene-phenotype information*

All different symptoms of CRC were extracted from HSDN and they are shown in Table 3. For each CRC symptom, genes causing similar phenotype were found through PhenomicDB [26]. Again, the similar scoring process which was used in the two previous steps was repeated for scoring genes based on gene-phenotype information.

Table 3. Human phenotypes related to CRC (extracted from HSDN).

| Symptoms for CRC | Score |
|---|---|
| Obesity | 81.59026 |
| Cachexia | 29.92086 |
| Weight Loss | 29.00156 |
| Body Weight | 18.17836 |
| Weight Gain | 10.40683 |
| Fever | 7.427252 |
| Overweight | 7.414849 |
| Asthenia | 6.189658 |
| Birth Weight | 5.628524 |
| Fetal Macrosomia | 3.336718 |

As we mentioned before, there were six different scoring schemes utilized in HybridRanker. To prioritize genes based on these six scores, a simple summation of them was used.

### 2.2. *Data extraction*

We used the human PPI network created by Lage et al. [27] which is an undirected weighted network. Interactions are extracted from different large scale data sources and their weights are based on network topology and experimental evidences. For noise reduction within the network, a cutoff threshold of 0.154 was defined for interaction weights. The filtered network contained 12,884 nodes (proteins) and 428429 edges (interactions).

CRC disease genes were extracted from DisGeNET which contains experimentally validated disease genes. By mapping CRC disease genes to the PPI network, 1121 genes in the network were marked as seed genes.

We explored MalaCards database for comorbid diseases and used HSDN to find diseases with similar symptoms. Also, genes with similar phenotype were extracted from PhenomicDB database.

### 2.3. *Comparison to other methods*

Leave-one-out cross-validation analysis was used to evaluate the HybridRanker and then its performance was compared with the performance of some other prioritization methods. Endeavour [28], ToppGene [4] and DIR [5] were three prioritization methods which their performances were compared with that of HybridRanker.

In each validation step, one seed gene was selected as a target gene and removed from the seed gene set. Then, we made an artificial linkage interval (containing 100 genes) with this target gene and its 99 nearest chromosomal neighbors (the chromosomal neighbors were obtained from UCSC database [29]). The remaining seed genes made a new seed gene set and the HybridRanker was applied on it. So, all the genes in the artificial linkage interval were given ranks based on the new seed gene set. The ranking of the target gene amongst other genes in the artificial linkage interval was the criterion for the performance evaluation. In other words, the higher rank the target gene is obtained, the better performance the algorithm has [12]. To show the performance of our method, receiver operating characteristic (ROC) curve was drawn to plot sensitivity versus 1-specificity.

Also, for better evaluation, the area under curve (AUC) was calculated for each curve. Here, the specificity was defined by the percentage of target genes (selected from the seed genes) ranked below a specific threshold and the sensitivity was the percentage of target genes which were ranked above that threshold.

For further evaluation of the HybridRanker, mean reciprocal rank (MRR) was used which is defined by:

$$\text{MRR} = \frac{1}{|Q|}\sum_{i=1}^{|Q|}\frac{1}{\text{rank}_i} \qquad (2)$$

where $Q$ shows the number of artificial linkage intervals and $rank_i$ was the ranking of $i^{th}$ removed seed gene in its related interval.

Also, we defined an average rank criterion which was calculated by averaging the ranks of all seeds in their corresponding intervals. The lower average rank shows the better performance of the algorithm. Also, the notations 1% and 5% were used to show the percentage of seed genes those were respectively ranked in the top 1% and 5% of their related intervals. All aforementioned criteria were computed for evaluating different methods.

### 2.4. *Gene Set Enrichment Analysis*

To examine the biological relevance of our result genes and CRC, we gave the top 100 ranked candidate genes as input for ClueGO, a Cytoscape plug-in, to integrate GO terms and pathway annotations. For pathway enrichment analysis, the KEGG, Wikipathways and Reactome databases which have been embedded in ClueGO plug-in were employed.

### 3. Results

We utilized a PPI network, with 12894 gens (proteins) in which 1121 nodes were labeled as seed genes based on experimentally examined causing genes of CRC. The SP algorithm (a method based on network local information), RWR and NP algorithms (methods based on network global information) were used to calculate the proximity of other nodes to the seed nodes. We used these three algorithms simoultansly to benefit both network local and global information. As it is shown in Figure1, the combination of network-based algorithms with biological knowledge (which resulted in the HybridRanker) had better perfomance than using only these three network-based algorithms.

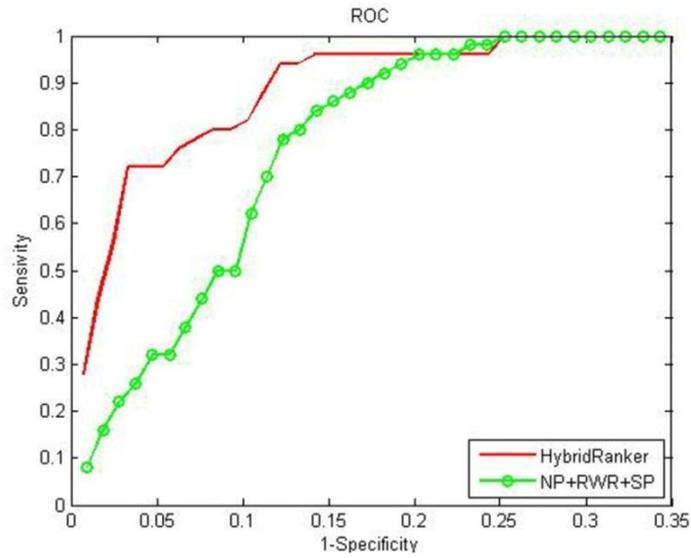

Figure 1. The NP, RWR and SP algorithms were applied on the PPI network and a combination of their scores was assigned to each gene. This figure shows the ROC curves for this combination and HybridRanker. In HybridRanker aforementioned algorithms were also composed with disease knowledge [23, 25, 26].

Also, for each network-based algorithm and each biological knowledge used in the HybridRanker, we evaluated their ability to prioritize genes. The results showed that neither of them is as powerful as the HybridRanker (see Figure 2).

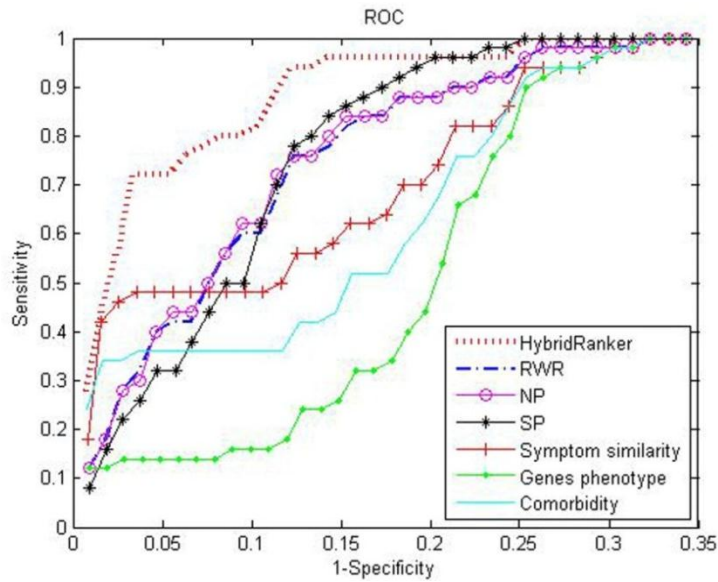

Figure 2. The ROC curves for HybridRanker and other scorings which are used in it.

We compared our proposed method with well known tools such as DIR, Endeavour and ToppGene which each has utilitized different data sources to prioritize genes. The ROC curves were drawn for each method (shown in Figure 3) and their AUC were also calculated (shown in Table 4). As it is shown in Table 4, the HybridRanker resulted in AUC of 0.95 which had the best performance among all examined methods. The NP and DIR methods were ranked as the second best methods with AUC of 0.91. Also, the average rank (See *comparison to other mthods* section) of the HybridRanker was 5.28 while for DIR, NP, RWR, SP, Endeavour and ToppGene it was 9.6, 9.8, 10, 10.1, 10.5 and 10.94, respectively. Table 4 also shows that in leave-one-out cross-validation analysis, 28% of the seed genes were achieved the highest ranking (top 1%) among genes in their related artificial linkage interval by using the HybridRanker which in comparison to other methods is the highest percentage. In addition, HybridRanker has the greatest percentage of the seed genes those are ranked in top 5% among genes in their related intervals. Furthermore, the results in Table 4 show that our method had the largest MMR value among all methods. Overall, all used criteria showed the better performanc of HybridRanker comparing to other examined algorithms.

Table 4. Different criteria for evaluating different disease gene prioritization methods.

| Methods | AUC | MRR | Average rank | 1% | 5% |
|---|---|---|---|---|---|
| **HybridRanker** | 0.95 | 0.46 | 5.28 | 0.28 | 0.72 |
| **Endeavour** | 0.76 | 0.25 | 10.5 | 0.10 | 0.40 |
| **ToppGene** | 0.70 | 0.28 | 10.94 | 0.14 | 0.34 |
| **DIR** | 0.91 | 0.28 | 9.6 | 0.18 | 0.38 |
| **RWR** | 0.90 | 0.25 | 10 | 0.12 | 0.40 |
| **SP** | 0.90 | 0.14 | 10.1 | 0 | 0.18 |
| **NP** | 0.91 | 0.25 | 9.8 | 0.12 | 0.40 |

Additionally, giving the top 100 ranked candidate genes to ClueGO showed that these genes were significantly enriched with GO terms and pathways most of which are consistent with several studies in CRC (see Table 5). Also, to examine the relevance of the obtained GO terms and CRC, we utilized the Comparative Toxicogenomics Database (CTD) [30]. Inferred associations between GO terms and diseases are provided in the CTD. We found that among GO terms listed in Table 5, four terms included negative regulation of DNA replication, response to UV, ovulation cycle process and stem cell proliferation are associated with CRC.

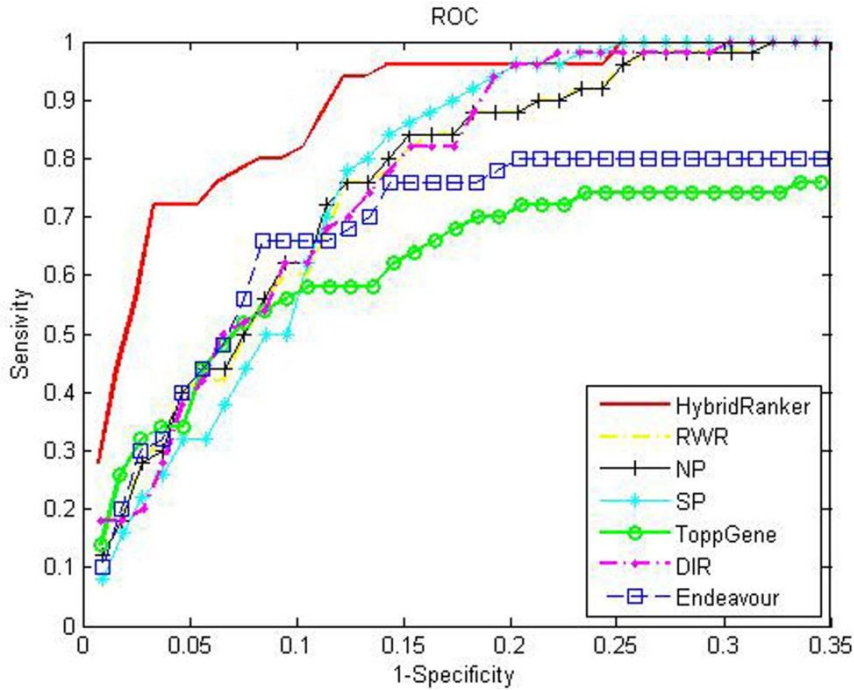

Figure 3. The ROC curves for HybridRanker and all other examined ones.

We also investigate the interactions between the top 100 ranked candidate genes and the seed genes. For each top gene, the lengths of shortest paths to all seed genes were computed and the minimum value of these lengths was considered. The results showed that 96% of these top genes were at the distance at most two from a seed gene (83% of them had direct interactions). Also, we found out that there were 3638 genes that participated in the shortest paths between the top genes and the seed ones. Among these intermediate genes, ten genes participated in the shortest paths between top candidates and seed genes, for more than a thousand times (listed in Table 6). In addition, the results showed that 99% (1112 out of 1121) of the seed genes had direct interactions to the top ranked candidate genes.

For further analysis, we examined the expressions of the top 100 ranked candidate genes in an expression dataset of CRC samples. The dataset used in this paper was obtained from Gene Expression Omnibus (GEO) database [43] at accession number GSE32323 (with 17 pairs of cancer and non-cancerous tissues from CRC patients). We used Mann-

Whitney test to identify genes that were differentially expressed. The results showed that 46 of the top candidate genes were differentially expressed between normal and tumor tissues (p-value<0.05).

**Table 5.** Related Go terms and pathways to the top 100 ranked candidate genes in HybridRanker which most of them are consistent with several studies in CRC.

| GO/Pathway ID | GO/Pathway Term | Paper |
|---|---|---|
| WP:24 | Peptide GPCRs | [31] |
| GO:0000245 | spliceosome complex assembly | [32] |
| GO:0007189 | adenylate cyclase-activating G-protein coupled receptor signaling pathway | [33] |
| GO:0008156 | negative regulation of DNA replication | [34] |
| GO:0008589 | regulation of smoothened signaling pathway | [35] |
| GO:0009411 | response to UV | [36] |
| GO:0010107 | potassium ion import | [37] |
| GO:0022602 | ovulation cycle process | |
| GO:0034367 | macromolecular complex remodeling | [38] |
| GO:0045834 | positive regulation of lipid metabolic process | [39] |
| GO:0045913 | positive regulation of carbohydrate metabolic process | [40] |
| GO:0072089 | stem cell proliferation | [41] |
| GO:1904036 | negative regulation of epithelial cell apoptotic process | [42] |

**Table 6.** The genes which are mediate in shortest paths between the top 100 ranked candidate genes of HybridRanker and seed genes for more than 1000 times.

| Gene symbol |
|---|
| PYGL |
| RNPEP |
| CMA1 |
| PLCXD3 |
| TRABD |
| GSR |
| BARHL2 |
| ORC5 |
| TRPM1 |
| HINT1 |

## 4. Discussion

In this paper, HybridRanker method is proposed as a new prioritization method to overcome disadvantages of pure network-based methods. The HybridRanker integrated information about disease similarity, disease comorbidity and phenotype similarity with human PPI network to utilize the advantages of both biological knowledge and network topological features. Similar phenotypes are assumed to be associated with the same causing genes [44, 45]. Based on that, we exploited two kinds of data in scoring new candidate disease genes: (1) known causing genes of diseases with similar symptoms to CRC and (2) genes related to the phenotypes similar to the CRC phenotypes. Also, disease comorbidity gives another important information which has been rarely considered in gene prioritization studies. Besides, there is an accepted assumption that disease proteins tend to have interactions with each other [46]. So, in HybridRanker we combined all these knowledge to gain better results in cancer gene prioritization.

**Conflict of interest**

The authors declare that they have no conflict of interest.